\documentclass[conference]{IEEEtran}
\IEEEoverridecommandlockouts
\usepackage{cite}
\usepackage{amsmath,amssymb,amsfonts}
\usepackage{algorithmic}
\usepackage{graphicx}
\usepackage{textcomp}
\usepackage{xcolor}
\usepackage{subfigure}
\usepackage{url}
\def\BibTeX{{\rm B\kern-.05em{\sc i\kern-.025em b}\kern-.08em
    T\kern-.1667em\lower.7ex\hbox{E}\kern-.125emX}}

\begin{document}

\title{BBR Fairness Evaluation Using NS-3\\
}

\author{\IEEEauthorblockN{Linchuan Tang}
\IEEEauthorblockA{\textit{Georgetown University} \\
lt665@georgetown.edu}
}

\maketitle

\begin{abstract}
This paper evaluates the fairness of BBR congestion control using NS-3 simulator. While BBR improves performance over loss-based methods in single flows, unfairness issues emerge with competing BBR and BBR/Cubic flows. Unfairness correlates with factors like round-trip time and buffer size. The core reason is the lack of responding mechanisms for the flows to converge on fair bandwidth share.  
\end{abstract}


\section{Introduction}
Loss-based congestion control mechanisms, such as Cubic, rely on detecting packet loss as a signal of network congestion. However, this suffers from a well-known issue called buffer bloat. It occurs when the sender put excessive packets beyond the buffer capacity of intermediate network devices. This allows the sender to detect packet loss and indirectly measure current congestion, but significantly increases the queuing delay and the round-trip time (RTT). Moreover, as advances in network interface controller (NIC) have enabled a higher throughput, the memory price goes down, loss-based methods show less flexibility in cooperating diverse network setting \cite{cardwell_cheng_gunn_yeganeh_jacobson_2016}.

In 2016, Google introduced an essentially different method, BBR: Congestion-Based Congestion Control \cite{cardwell_cheng_gunn_yeganeh_jacobson_2016}. BBR was designed to address the two issues and improve throughput. It drew a wide attention that its strengths have been extensively verified, various shortcomings have been discovered. This study mainly examines the characteristic of BBR and the fairness issues when BBR interplays with loss-based congestion control. 

Using NS-3 network simulator, this study validates the improvement in reducing the delay and queuing delay. However, competing simulations show unfairness issue correlating with different bottleneck queue size or round-trip time (RTT). Finally, we identify the root cause is the lack of convergence response between different controls.

Source code and results are available at \url{https://github.com/tljk/NS3-bbr}

\section{Related Work}

Traditional TCP congestion controls can be classified according to the method of detecting congestion, mainly divided into two categories, loss-based control and delay-based control. Loss-based control, like the widely deployed BIC \cite{xu2004binary}, TCP Cubic \cite{ha2008cubic}, primarily detect packet loss as an indicator of congestion. On the other hand, delay-based control, such as TCP Vegas \cite{brakmo1994tcp}, focus on monitoring the round-trip time (RTT), interpreting an increasing delay as a signal of congestion. However, delay-based approaches have not seen widespread adoption in industry. 

BBR has gained significant attention from both academia and industry. While its behavior and the performance are widely validated in single flow scenario. Multiple issues emerge in multi-flow contexts. Hock et al. \cite{hock_bless_zitterbart_2017} observed multiple BBR flows overestimating bandwidth in a shared bottleneck, leading to excessive delay and buffer. They also highlighted unfair share and packet loss problem in shallow buffer. Miyazawa et al. \cite{miyazawa_sasaki_oda_yamaguchi_2018} discovered drastic periodic competitions between BBR and Cubic. They drew similar conclusions but explained it as deficiencies in BBR’s bandwidth estimation. Scholz et al. \cite{scholz_jaeger_schwaighofer_raumer_geyer_carle_2018} introduced a TCP measurement framework for replicating and thoroughly analyzing previous work. They further investigated in the synchronization mechanism and identified shortcomings with short-term flows. Ware et al. \cite{ware2019modeling} took a modeling and verification approach to analyze multi-flow interaction. Showing that BBR’s unfairness arises as its bandwidth estimations exclude the shares of competing loss-based controls. Claypool et al. \cite{claypool2019sharing} verified prior findings and suggested adding feedback loops to improve responsiveness to network changes. Zhang et al. \cite{zhang_zhu_xia_zhang_zhang_deng_2019} conducted NS-3 simulations, attributing fairness issue to the mechanism difference in adjusting congestion window. They also discussed the combined effect of BBR and different active queue disc. More recent works have focused on developing BBR variants in improving fairness or adapting to various network environment.

In summary, issues with BBR include:
\begin{itemize}
    \item Unfairness over different RTT
    \item Bandwidth overestimate on multi-flow
    \item High packet retransmit rate in shallow buffer
    \item Unfair bandwidth share depending on buffer size
    \item Performance degradation of coexisting Cubic flows
\end{itemize}

\section{BBR Overview}
The performance of a TCP connection primarily depends on the attributes of the bottleneck link. The data rate of the bottleneck link determines the overall throughput of the entire connection. Meanwhile, increasing queuing delays are attributed to bufferbloat at the bottleneck. The connection operates at its best when round-trip time (RTT) closely matches the round-trip propagation delay of the link, sending rate closely aligns with available bottleneck bandwidth. This is the congestion model that motivates the design of BBR's mechanisms:
\begin{equation}
BDP=BtlBw\times RTprop\label{eq1}
\end{equation}
where BDP stands for bandwidth delay product, BtlBw for bottleneck bandwidth, and RTprop for round-trip propagation time. BDP represents the optimum amount of inflight for this connection. By operating around BDP, the connection can provide the minimum delay and maximum throughput.

BBR is a congestion control that is designed to estimate and adapt to the dynamic nature of bottleneck bandwidth and round-trip delay. Its primary performance objectives include achieving a higher throughput with minimized delay. Moreover, it further targets to achieve a fairness share of bandwidth for both BBR and non-BBR flows.

BBR’s operation can be described in the following three states.

\subsection{Start-up}
Similar to TCP slow start, BBR exponentially increases the sending rate with a pacing gain of 2/ln2  to probe available capacity. It stops when delivery rate fails to substantially grow (less than 1.25 times) over three RTTs.

\subsection{Drain}
Upon filling queues, BBR enters drain state to clear the excessive packets in the queue, which may accumulate up to 3 BDP. It uses an inverse of startup pacing gain, persisting until inflight drops to one BDP.

\subsection{Steady}
In steady state, BBR periodically probes for bandwidth change. It increases the sending rate by setting a 25\% pacing above the current estimated rate for one RTT and measures whether deliver rate increases. If no change in bandwidth, it compensates by decreasing the pacing 25\% below the estimated rate for one RTT. Within each cycle, BBR also checks the change of RTT. If the minimal value of RTT has not been updated over the past 10 seconds, BBR drains the queue by setting the inflight to only 4 packets, and thereby estimates a new RTprop accordingly. Additionally, it may re-enter startup state when the flow is not in full capacity.

BBR takes advantage of those two estimations and make adjustment of congestion window size (CWND) and pacing rate. Congestion window will match the size of a single BDP. Pacing rate will align with the estimated bottleneck rate. 

\section{Experiment Setup}
\begin{figure}[htbp]
\centerline{\includegraphics[width=1\linewidth]{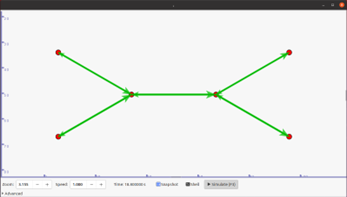}}
\caption{NS-3 dumbbell visualization.}
\label{dumbbell_vis}
\end{figure}

Ns-3 is an open-source network simulator widely used in education and research. Similar to other experiment or evaluation over TCP BBR \cite{hock_bless_zitterbart_2017,scholz_jaeger_schwaighofer_raumer_geyer_carle_2018,zhang_zhu_xia_zhang_zhang_deng_2019}, we employ dumbbell topology  (Fig.~\ref{dumbbell_vis}), in which two gateway switches connected by a single bottleneck link. A switch can connect to multiple end devices, which act as senders or receivers. Dumbbell topology simulates a congestion network that is characterized by a single bottleneck link. It enables analysis of interaction between congestion control and dynamics on bottleneck link. This topology can also be adapted to a point-to-point network for individual method.

The bottleneck link characteristics include bandwidth, delay, and buffer size configured. The experiment then evaluates congestion control under single and multiple flow conditions.

\begin{table}[htbp]
\caption{Default configuration of dumbbell topology}
\begin{center}
\begin{tabular}{|l|l|} \hline  Bottleneck link rate&10 Mbps\\ \hline 
 Leaf link rate&100 Mbps\\ \hline 
 Propagation delay&100 ms\\ \hline 
 Queue size&1.5 BDP\\ \hline 
 Packet loss rate&0 \%\\ \hline\end{tabular}
\label{tabel_set}
\end{center}
\end{table}
Table \ref{tabel_set} gives the default configuration for our experiment. We primarily use three kinds of dumbbell setups.
\begin{itemize}
    \item Point-to-point single flow experiments
    \item Multiple BBR flows experiments
    \item Multiple BBR, Cubic flows experiments
\end{itemize}
Our experiment results are derived from NS-3 tracing system. It tracks the change of event and data that we are interested in, which includes
\begin{enumerate}
    \item Throughput, link utilization in megabytes per second. In average throughput (in log.txt) and throughput measurement every 0.1 second.
    \item  Delay, round-trip time (RTT) measurement for each packet. We also include the retransmission timeout (RTO).
    \item Queue size, the buffer load of the router at bottleneck for every 0.1 second.
    \item Inflight, the amount of packets being transmitted.
    \item Congestion window size (CWND).
    \item Packet loss, the number of lost packet for every 0.1 second.

\end{enumerate}

\section{Evaluation}
\subsection{Single Flow Scenario}
\begin{figure}[htbp]
\centering
\subfigure[Throughput comparison between BBR and Cubic. Average throughput for Cubic: 9.12 Mbps, for Cubic: 9.31 Mbps]
{\includegraphics[width=1\linewidth]{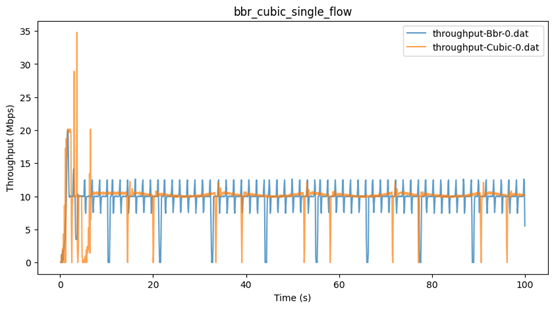}
\label{single_flow.a}}
\subfigure[Round-trip time comparison between BBR and Cubic]
{\includegraphics[width=1\linewidth]{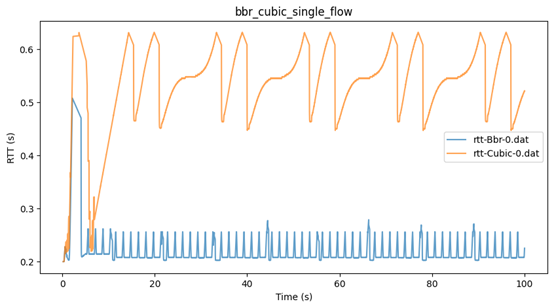}
\label{single_flow.b}}
\subfigure[Bottleneck link queue length comparison between BBR and Cubic]
{\includegraphics[width=1\linewidth]{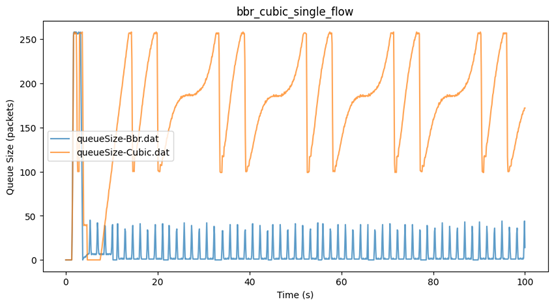}
\label{single_flow.c}}
\caption{Single flow experiment at default configuration}
\label{single_flow}
\end{figure}
The single flow experiment compares the individual behavior of BBR and Cubic and the divergences between two congestion controls. 

Initially, both methods try to exhaust the link capacity (Fig. \ref{single_flow.a}). While the purpose of BBR is to quickly estimate the bottleneck bandwidth. As visualized in Fig. \ref{single_flow.c}, BBR then enters drain state to clear the excessive packets on bottleneck and nearly operates at no queue. By comparison, Cubic persists in maintaining a high queue load. 

Correspondingly, in Fig. \ref{single_flow.b}, BBR has a significant lower delay approaching the underlying propagation delay, whereas Cubic reflects persistent  bufferbloat.

Besides, BBR maintains two probing cycles. Every 10 seconds, BBR performs a round-trip time (RTprop) estimation as its window expires. This triggers draining the bottleneck buffer to capture the underlying propagation delay, by reducing the congestion window to only 4 packets. It corresponds to the periodic steep drops in throughput and congestion window for every 10 seconds.

Another is bandwidth probe loops, which is performed in a cycle of 8 RTTs. In the first RTT, BBR makes a pacing increase by 25\%. It then compensates for another RTT by decreasing the pacing by 25\%. For the remaining 6 RTTs, pacing holds steady. For every 10 seconds, BBR gives about 6 cycles of bandwidth probing, since we have a 0.2 RTT in our experiments. This will correspond to 6 small spike-like increases in throughput, congestion window, RTT and queue length measurements.

Overall, in this loss-free scenario, both BBR and Cubic achieve a high utilization (over 90\%) and a slightly higher result on Cubic. However, BBR maintains a significant lower RTT and queue length without losing throughput.

\begin{table}[htbp]
\caption{Single flow experiment with different error rate}
\begin{center}
\begin{tabular}{|l|l|l|}\hline
 Error Rate& \multicolumn{2}{|c|}{Throughput (Mbps)}\\\hline \hline  &BBR &Cubic\\ \hline 
 0.0001&9.12&9.22\\ \hline 
 0.001&9.08&3.93\\ \hline 
 0.01&1.09&0.92\\\hline\end{tabular}
\label{tabel_err}
\end{center}
\end{table}
We also introduce a random packet drops at given rate (Table \ref{tabel_err}) to simulate operation of congestion control under extremely poor network conditions. 

As a loss-based mechanism, the performance of Cubic is intrinsically sensitive to loss rate. We observe such performance downgrade when increasing the error rate. On the contrary, BBR is designed not to respond to packet loss and has a much higher tolerance of error. 

Original BBR evaluations found negligible throughput impact below an error rate of 5\% and a significant degradation over an error rate of 10\% \cite{cardwell_cheng_gunn_yeganeh_jacobson_2016}. Our simulation largely confirms BBR’s resilience to about 1\% loss. Throughput persists near capacity over the initial 20 seconds, though retransmissions and inflating RTTs gradually accumulate and causing performance degradation. This is not an intended behavior. A similar issue was reported in \cite{RN16}.

\subsection{Multiple BBR Flows}
Multiple BBR flows experiment shows the distributed convergence behavior. It includes two key aspects of convergence: fair bandwidth allocation and the aligned RTprop estimation.

\subsubsection{Assess BBR Fair Share}
\begin{figure}[htbp]
\centering
\subfigure[Throughput measurement of two BBR flows at default configuration]
{\includegraphics[width=1\linewidth]{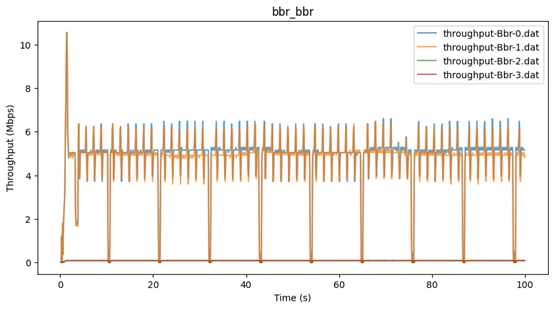}\label{sync.a}}
\subfigure[Congestion window measurement of two BBR flows with 10ms delay]
{\includegraphics[width=1\linewidth]{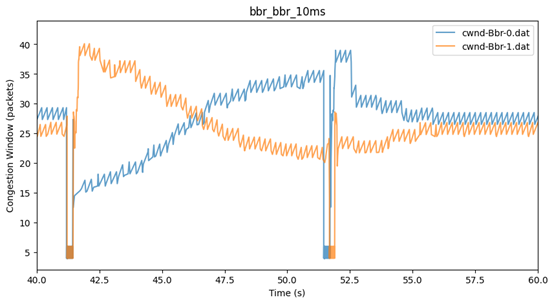}\label{sync.b}}
\subfigure[RTT measurement of two BBR flows with 10ms delay]
{\includegraphics[width=1\linewidth]{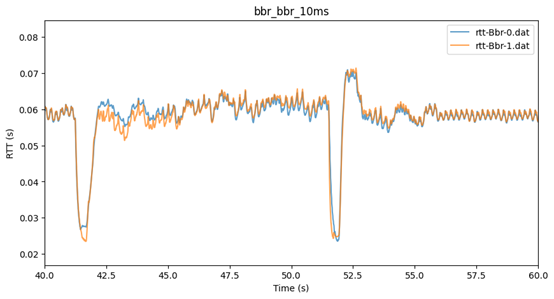}\label{sync.c}}
\caption{Synchronization behavior between two BBR flows}
\label{sync}
\end{figure}
Instead of probing bottleneck capacity, BBR probes the bandwidth share in multi-flow scenario. The probing methodology effectively redistributes bandwidth from over-utilized flows to under-utilized ones. This is obtained from BBR's interpretation of increasing delivery rates and RTTs measured.

\begin{equation}
deliver\ rate=packet\ size / RTT\label{eq2}
\end{equation}
If there is no bandwidth change, increasing sending rate can only create excessive queue on bottleneck. Although there are more packets being transmitted, the increase in throughput will be offset by continually increasing RTTs, and there will be no improvement in deliver rate. Conversely, if an increase in sending rate does give an increase in deliver rate. From equation \ref{eq2}, it suggests that RTT is not fluctuating as significant as previous probe. The fluctuation drop gives a higher measure of deliver rate and indicates a spare capacity. Meanwhile, a further increase in sending rate may induce a stronger RTT fluctuation with no deliver rate improvement. This indicates an overestimate of available bandwidth and the sending rate. BBR flow uses these RTT fluctuations to determine its relative standing.

This mechanism extends to fair sharing across multiple flows. A flow occupying a high bandwidth share will hold greater inflight, contributes a higher share of queue growth when probing the bandwidth. It gives a higher variation in RTT measurement. In contrast, flows with low share experience smaller RTT variations from probing. As a result, flow with a high share observes RTT inflation and decreases its sending rate. While flow with a low share see more capacity and increase their share. Through repeated iterations, the probing outcomes synchronize. All flows exhibit equivalent RTT behaviors in response to pacing increase. It demonstrates that all flows converge in the same estimation of their available bandwidth, and thus an agreement on fair share is reached.

Fig. \ref{sync} give an example of how multiple BBR flows reach a fair share. At approximately 42 seconds, both flows enter an RTprop probing state, decreasing their congestion window. When both flows explore the bandwidth, flow 1 gains a higher share and compromises flow 0. The RTT response reveals how to restore equity. Flow 1 first demonstrates a higher magnitude of inflation from its excessive contribution to queuing. It signals an overestimate in bandwidth, resulting in a decrease in sending rate. Meanwhile, flow 0 continues to gain the spare share and increase its sending rate. Two flows oscillate towards a fair share, eventually synchronize, showing parallel bandwidth estimations and similar profiles in RTT fluctuation.

\subsubsection{RTT Synchronization}

Additionally, no single flow can accurately sample the true minimum delay when other traffic presents. BBR requires alignment of RTprop estimations, which ensures all flows can accurately sample the minimum delay. When a dominant flow enters probe RTprop state, it drains the buffer, which gives a lower RTT measurement for other flows. By observing a new measurement, all other flows enter the probing stage and drain the entire connection. As a result, multiple flows reach an alignment that each flow enter probing state simultaneously and make new estimation without interference.

\subsubsection{Bandwidth Overestimate}

Another well observed behavior is the bandwidth overestimate with multiple BBR flows. When flows share a bottleneck, each individually gives a slightly higher estimation on their available bandwidth. Our experiments measure the approximately stable congestion window size at 90 seconds to quantify this effect.

\begin{table}
\centering
\caption{Total Congestion window size of BBR flows}
\label{table_over}
\begin{tabular}{| l | l | l | l | l | l |}
\hline
  & 1 flow & 2 flows & 3 flows & 5 flows & 8 flows \\
\hline
Cwnd (packets) & 342 & 360 & 395 & 415 & 444 \\
\hline
Cwnd increase& 0 & 0.05 & 0.10 & 0.21 & 0.30 \\
\hline

\end{tabular}

\end{table}
As Hock et al. \cite{hock_bless_zitterbart_2017} described, if the bandwidth probe imperfect aligned, each flow will aggressively increase their share of bandwidth. This overestimate induces issues like bufferbloat, and packet loss in shadow buffer. Our experiments use congestion window size to approximate the estimate bandwidth for each flow. Our results (Table \ref{table_over}) suggest that add one flow can induce a 5\% of overestimate, but total inflation will not exceed 30\%. This aligns with the upper limit of overestimate that they suggested.

\begin{figure}[htbp]
\centering
\subfigure[Throughput measurement between BBR flow and Cubic flow]
{\includegraphics[width=1\linewidth]{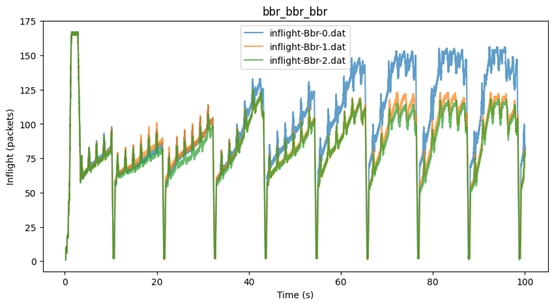}\label{overestimate.a}}
\subfigure[RTT measurement for three BBR flows]
{\includegraphics[width=1\linewidth]{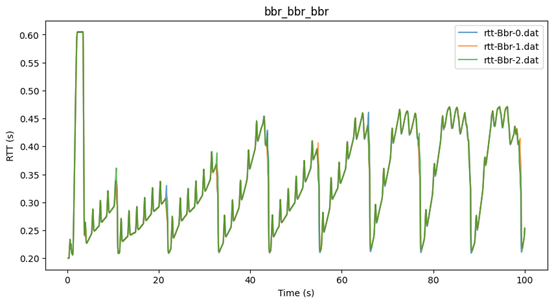}\label{overestimate.b}}
\subfigure[Bottleneck queue length for three BBR flows]
{\includegraphics[width=1\linewidth]{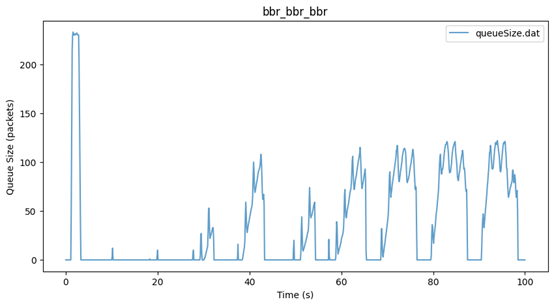}\label{overestimate.c}}
\caption{Overestimate bandwidth behavior for three BBR flows at default configuration}
\label{overestimate}
\end{figure}
Fig. \ref{overestimate.a} shows the collective growth of inflight when multiple flows coexist. The measurement of congestion window suggest that each flow gives a slight overestimate and induce the increase in inflight. The excessive packets cause the increase in queue length and RTT, in Fig. \ref{overestimate.b}\ref{overestimate.c}.

\subsection{BBR/Cubic Flows}
Multiple BBR, Cubic flows experiment shows the coexisting behavior of multiple BBR flows and Cubic flows. We mainly experiment with one BBR flow and one Cubic flows. In our experiments, we observe that the unfair behavior is close related to the settings of bottleneck queue size and propagation delay.

\subsubsection{RTT Unfairness}

Experiments with different RTT settings \cite{scholz_jaeger_schwaighofer_raumer_geyer_carle_2018} shows BBR’s biases in bandwidth allocation. BBR flow with a high RTprop obtains a high share of the bandwidth, while low RTprop flow gets a low share. This attributes to two reasons.

First, BDP estimation is directly proportional to RTprop. A flow with a higher delay inherently targets at a greater volume.

Second, BBR’s pacing gain cycle length stays at one RTT. Flows with a higher RTT thus probe and transmit at an extended timescale compared to flows with smaller RTT. This enables them to flood the buffer with extra packets and compromises the share of others.

\subsubsection{Buffer Size Effects}

Unfairness issues also relate to bottleneck queue size. BBR generally puts about 1 BDP of inflight and up to 1.5 BDP when probing bandwidth. Prior work by Claypool et al. \cite{claypool2019sharing} suggests that the decrease in queue size to below 1.5 BDP gives an increase in packet loss. We replicate the behavior and observe a serious performance downgrade in shallow buffer. Accordingly, initial experiment starts with a 1.5 BDP (optimal point) of queue size with a 100ms propagation delay. Subsequent experiments scale queue size down to 0.8 BDP and up to 3, 6 BDP while reducing propagation delay to 10ms and 50ms.

\begin{figure}[htbp]
\centering
\subfigure[Throughput measurement between BBR flow and Cubic flow at 1.5 BDP]
{\includegraphics[width=1\linewidth]{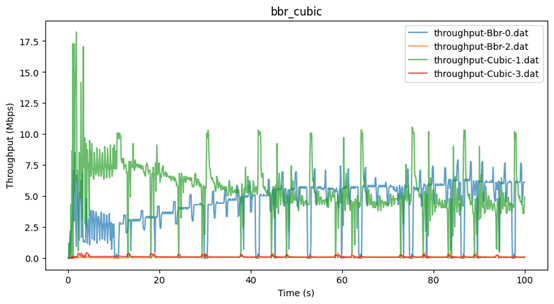}\label{comp.a}}
\subfigure[Throughput measurement between BBR flow and Cubic flow at 0.8 BDP]
{\includegraphics[width=1\linewidth]{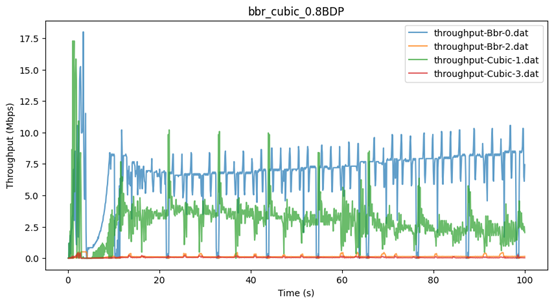}\label{comp.b}}
\subfigure[Throughput measurement between BBR flow and Cubic flow at 3 BDP]
{\includegraphics[width=1\linewidth]{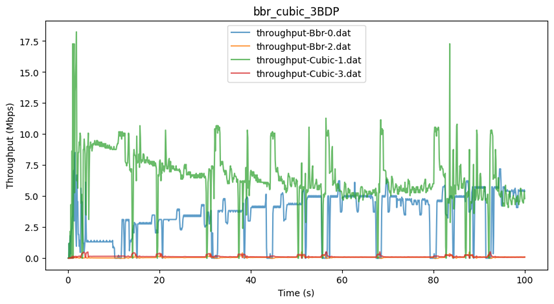}\label{comp.c}}
\subfigure[Throughput measurement between BBR flow and Cubic flow at 6 BDP]
{\includegraphics[width=1\linewidth]{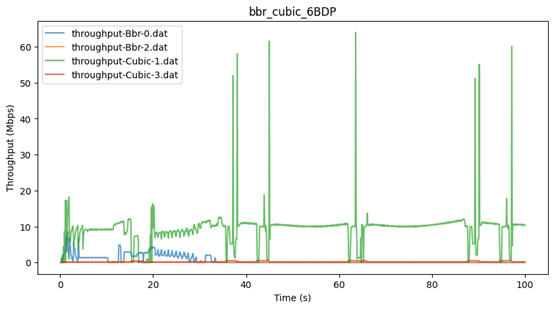}\label{comp.d}}
\caption{Throughput measurement between BBR flow and Cubic flow}
\label{comp}
\end{figure}
With a queue size equals or larger than 6 BDP (Fig. \ref{comp.d}), Cubic dominates. It actively exhausts the queue and increases the overall delay. BBR accordingly reduces sending rate for each cycle, while Cubic further increases its share and eventually take over the entire bandwidth.

At 3 BDP (Fig. \ref{comp.c}), periodic oscillations occur, achieving an approximate fairness. By putting excessive packets to queue, Cubic occupies advantage at the beginning of each cycle. However, BBR tit-for-tat further increases its congestion window and gain more share from Cubic. This is achieved through the aggressive bandwidth probe, where BBR increase its sending rate by 1.25 times. This compromises the share of Cubic by overwhelming the queue and inducing back-off. It yields a temporary equity, though exact convergence remains elusive.

In Fig. \ref{comp.c}, each time BBR performs a bandwidth probe, Cubic’s share compromises. However, similarly, when BBR enters RTT probing stage, Cubic aggressively gains bandwidth share and dominates at the start of the next cycle. 

If we further reduce the queue size to 1.5 (Fig. \ref{comp.a}) or 0.8 (Fig. \ref{comp.b}) BDP, BBR dominates, and its bandwidth share increases. Bandwidth probing still provides the same amount of excess packets. However, as we reduce the queue size, it still increases proportionally. Experiments show that relatively more packets are put into the queue, inducing more frequent back off. This demonstrates why there is more severe retransmission and packet loss in a shallow buffer.

\begin{figure}[htbp]
\centering
\subfigure[Throughput measurement between BBR flow and Cubic flow]
{\includegraphics[width=1\linewidth]{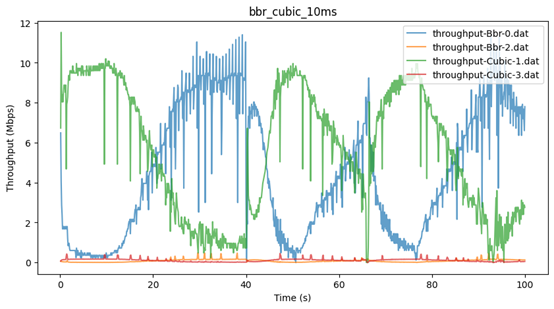}\label{comp10.a}}
\subfigure[Congestion window size measurement between BBR flow and Cubic flow]
{\includegraphics[width=1\linewidth]{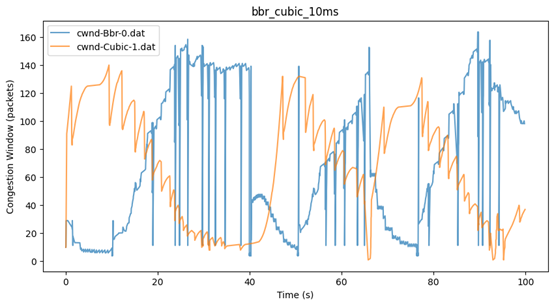}\label{comp10.b}}
\caption{Throughput measurement between BBR flow and Cubic flow}
\label{comp10}
\end{figure}
Miyazawa et al. \cite{miyazawa_sasaki_oda_yamaguchi_2018} characterized competitive cyclic behavior between BBR and Cubic flow with low propagation delays. A reduced RTTs shrink the pacing gain interval, enabling more frequent and disruptive probing. As depicted in Fig. \ref{comp10.a}, Cubic initially dominates at each cycle, However, it suffers from the frequent bandwidth probe from BBR, leading to frequent packet loss and back off. BBR interprets this continuously back off as available capacity. The relatively small congestion window and more frequent bandwidth gain make it eventually takes over the bottleneck. When BBR reduces its congestion window, Cubic rapidly regains high share and aggressively inflates buffer, which trigger BBR to reduce its bandwidth share. As BBR gradually retain its share, the cycle starts again.

Although BBR and Cubic can respond to each other’s signals, they fail to achieve harmonious sharing. In Fig. \ref{comp10.b}, BBR gives a fluctuating estimation of share when Cubic presents. In fact, fairness issues can be traced back to the instability or inequality in its bandwidth estimation. Among BBR flows, bandwidth probing act as a negative feedback loop that converges on fair bandwidth share. However, BBR competes against Cubic in a positive loop that either goes extreme or oscillation. There is no responding mechanism to achieve convergence between two methods. This is why there is rare fairness between BBR and Cubic.

\section{Conclusion}
Experiments validate BBR's single flow advantage but reveal deficiencies with competing flows. Unfairness happens when multiple BBR and BBR/Cubic flows share a bottleneck, which relates to different settings of round-trip times and buffer sizes. Essentially, the reason for the unstable or unequal unfairness is the absence of responding mechanisms for bandwidth convergence.  

Potential ways to improve fairness include, enhancing the accuracy and robustness of bandwidth estimation; introducing a mechanism for responding to signals from loss-based controls. 

\bibliographystyle{IEEEtran}
\bibliography{conference}
\vspace{12pt}

\end{document}